\pdfoutput=1

\documentclass[10pt,twocolumn,letterpaper]{article}

\usepackage{iccv}
\usepackage{times}
\usepackage{epsfig}
\usepackage{graphicx}
\usepackage{amsmath}
\usepackage{amssymb}

\usepackage{savesym}
\savesymbol{checkmark}

\usepackage{xcolor}		
\usepackage[colorlinks = true,
            linkcolor = purple,
            urlcolor  = blue,
            citecolor = cyan,
            anchorcolor = black]{hyperref}	

\usepackage[inline]{enumitem}
\usepackage{booktabs}
\usepackage{dingbat}
\usepackage{xcolor}
\usepackage{caption}
\usepackage{authblk}
\usepackage[toc,page]{appendix}
\usepackage{algorithm}
\usepackage{algorithmic}
\usepackage{subfig}
\iccvfinalcopy 

\usepackage{scalerel}
\usepackage{tikz}
\usetikzlibrary{svg.path}
\definecolor{orcidlogocol}{HTML}{A6CE39}
\tikzset{
  orcidlogo/.pic={
    \fill[orcidlogocol] svg{M256,128c0,70.7-57.3,128-128,128C57.3,256,0,198.7,0,128C0,57.3,57.3,0,128,0C198.7,0,256,57.3,256,128z};
    \fill[white] svg{M86.3,186.2H70.9V79.1h15.4v48.4V186.2z}
                 svg{M108.9,79.1h41.6c39.6,0,57,28.3,57,53.6c0,27.5-21.5,53.6-56.8,53.6h-41.8V79.1z M124.3,172.4h24.5c34.9,0,42.9-26.5,42.9-39.7c0-21.5-13.7-39.7-43.7-39.7h-23.7V172.4z}
                 svg{M88.7,56.8c0,5.5-4.5,10.1-10.1,10.1c-5.6,0-10.1-4.6-10.1-10.1c0-5.6,4.5-10.1,10.1-10.1C84.2,46.7,88.7,51.3,88.7,56.8z};
  }
}

\usepackage{tikz,xcolor,hyperref}

\definecolor{lime}{HTML}{A6CE39}
\DeclareRobustCommand{\orcidicon}{
	\begin{tikzpicture}
	\draw[lime, fill=lime] (0,0)
	circle [radius=0.16]
	node[white] {{\fontfamily{qag}\selectfont \tiny ID}};
	\draw[white, fill=white] (-0.0625,0.095)
	circle [radius=0.007];
	\end{tikzpicture}
	\hspace{-2mm}
}
\foreach \x in {A, ..., Z}{\expandafter\xdef\csname orcid\x\endcsname{\noexpand\href{https://orcid.org/\csname orcidauthor\x\endcsname}
			{\noexpand\orcidicon}}
}

\def\keywordname{{\bfseries \emph Keywords}}%
\def\keywords#1{\par\addvspace\medskipamount{\rightskip=0pt plus1cm
\def\and{\ifhmode\unskip\nobreak\fi\ $\cdot$
}\noindent\keywordname\enspace\ignorespaces#1\par}}

\ificcvfinal\fi
\begin{document}

\title{Ultrasound Phase Aberrated Point Spread Function Estimation with Convolutional Neural Network: Simulation Study}

\renewcommand*{\Authfont}{\bfseries}
\author[1]{Wei-Hsiang Shen\orcidA{}}
\author[1]{Yu-An Lin}
\author[1,2,3,*]{Meng-Lin Li\orcidB{}}

\affil[1]{Department of Electrical Engineering, National Tsing Hua University, Hsinchu, Taiwan}
\affil[2]{Institute of Photonics Technologies, National Tsing Hua University, Hsinchu, Taiwan}
\affil[3]{Brain Research Center, National Tsing Hua University, Hsinchu, Taiwan}
\affil[*]{Correspondence: \texttt{mlli@ee.nthu.edu.tw}}

\maketitle
\begin{abstract}
Ultrasound imaging systems rely on accurate point spread function (PSF) estimation to support advanced image quality enhancement techniques such as deconvolution and speckle reduction. Phase aberration, caused by sound speed inhomogeneity within biological tissue, is inevitable in ultrasound imaging. It distorts the PSF by increasing sidelobe level and introducing asymmetric amplitude, making PSF estimation under phase aberration highly challenging. In this work, we propose a deep learning framework for estimating phase-aberrated PSFs using U-Net and complex U-Net architectures, operating on RF and complex k-space data, respectively, with the latter demonstrating superior performance. Synthetic phase aberration data, generated using the near-field phase screen model, is employed to train the networks. We evaluate various loss functions and find that log-compressed B-mode perceptual loss achieves the best performance, accurately predicting both the mainlobe and near sidelobe regions of the PSF. Simulation results validate the effectiveness of our approach in estimating PSFs under varying levels of phase aberration.





\end{abstract}

\let\thefootnote\relax\footnotetext{This work is supported by National Science and Technology Council, Taiwan, under the grant number NSTC 113-2321-B-002-029- and MOST 110-2221-E-007-011-MY3.}

\section{Introduction}
Point spread function (PSF) represents the two-dimensional spatial impulse response of an ultrasound imaging system. It models and describes the characteristics of an ultrasound imaging system, such as resolution, contrast, and focus. Despite the maturity and widespread adoption of ultrasound imaging, ongoing efforts continue to enhance its image quality and clinical value through advances in techniques such as image deconvolution, speckle reduction, and localization microscopy \cite{review_SR,review_enhancement,review_us_postprocessing}. These techniques critically depend on accurate PSF estimation for optimal performance \cite{dalitz2015point,rangarajan2008ultrasonic}. For example, in image deconvolution, the PSF functions as the degradation kernel to be restored for recovering the original image. Similarly, in speckle reduction, the PSF may serve as a metric or guide to adjust the strength and direction of the reduction process.


Phase aberration is an inevitable and undesired side effect in ultrasound imaging due to the inhomogeneity of the speed of sound in human tissue \cite{flax1988phase,o1988phase,li2003adaptive,soulioti2021deconstruction,ali2023aberration}. It alters the amplitude and phase of the acoustic signal, degrading the focusing of ultrasound systems and resulting in reduced image contrast and quality. Moreover, it distorts the point spread function (PSF) of the imaging system \cite{anderson1997detection,yasuda2019phase}. The PSF after such distortion is transformed into a phase aberration version, typically exhibiting higher sidelobe energy and asymmetric amplitude. Accurate estimation of PSF with phase aberration remains a significant challenge, particularly for \textit{in vivo} imaging, due to its highly ill-posed nature. Estimating accurate PSFs in the presence of phase aberration continues to be an area of active research.

Prior methods for estimating the PSF in ultrasound imaging have utilized homomorphic filtering techniques to separate the PSF from the scatterer distribution, operating under the assumption that PSFs are spatially smoother compared to the randomly distributed scatterers \cite{kobayashi1984spectral}. Some strategies have enhanced the robustness of cepstrum estimation in homomorphic filtering by combining multiple one-dimensional cepstra using interpolation \cite{mattausch2016image}. Others have extended this approach by implementing phase unwrapping on the scatterer distribution within the cepstral domain \cite{taxt1995restoration,michailovich2005novel}. However, these methods are limited to well-focused PSFs and do not account for the distortion of PSFs caused by phase aberrations in \textit{in vivo} environments. Additionally, while some have proposed a parametric form to model the PSF in ultrasound imaging \cite{benameur2012homomorphic}, this parametric form, derived from a Gaussian distribution, also does not account for the effects of phase aberration.

Deep learning-based techniques have emerged as powerful tools in ultrasound imaging. Neural networks are being employed for a range of tasks, including lesion detection and segmentation \cite{noble2006ultrasound}, speckle reduction \cite{huang2020mimicknet}, image enhancement, and beamforming \cite{luchies2018deep}. A key advantage of deep learning lies in its data-driven approach; instead of relying on mathematical derivations, which can be overly complex or ill-posed, deep learning methods learn underlying relationships directly from the data.

Recent research employed a convolutional neural network (CNN) to estimate the aberration profile of the PSF from the received channel data \cite{sharifzadeh2020phase}. However, this approach requires re-beamforming to obtain the PSF using the predicted aberration profile. Additionally, operating on channel data presents a significant challenge in terms of memory efficiency when integrated into contemporary ultrasound scanners.


In this work, we introduce a novel deep learning technique for estimating phase-aberrated PSFs in ultrasound imaging. The method employs a convolutional neural network (CNN) that takes patches of beamformed speckle data as input and infers the underlying phase-aberrated PSFs. We evaluate two configurations: a U-Net architecture operating on radio frequency (RF) data and a complex U-Net architecture designed for complex k-space data. Estimating the PSF in the k-space simplifies the problem, as it resembles a denoising process and mitigates the influence of scatterers in the k-space representation. Our results demonstrate that the k-space approach outperforms the RF domain one in terms of estimation accuracy. The CNN is trained on synthetic phase-aberrated PSFs, leveraging a B-mode perceptual loss function to accurately measure PSF similarity and enhance the training effectiveness.



In summary, in this paper we present:

\begin{itemize}
\item A U-Net to infer the underlying RF phase-aberrated PSF from RF speckle data.
\item A complex U-Net operating in k-space for better performance in phase-aberrated PSF estimation.
\item A B-mode perceptual loss function that effectively quantifies PSF similarity, facilitating improved network training.
\item A detailed evaluation of different loss functions for training network to infer PSF.
\end{itemize}

The methods presented in the paper were first mentioned in \cite{9957752,yalin_revised_lrelu}, featuring simulation demonstration and networks trained in the spatial domain. This paper expands upon \cite{9957752,yalin_revised_lrelu} by introducing improved methods in the k-space, performing a comprehensive evaluation across various loss functions, and validating the approach with extensive simulation data.



\section{Materials and Methods}

\subsection{PSF Estimation as an Inverse Problem}

The ultrasound imaging system can be represented mathematically as a convolution model \cite{taxt1994radial}, as shown in Equation \ref{eqn:convolution_model}:

\begin{equation}
g(x,y)=f(x,y)*h(x,y)
\label{eqn:convolution_model}
\end{equation}

\noindent
where $g$ is the observed data, $f$ is the scatterer distribution, $h$ is the system’s underlying 2D PSF, and $*$ is the convolution operator.

Estimating the PSF presents an ill-posed inverse problem, where both the scatterer distribution $f$ and PSF $h$ are unknown, and the goal is to estimate the PSF $h$. In addressing this problem, we acknowledge the presence of phase aberration and aim to estimate the equivalent phase-aberrated PSF. In this work, we employ convolutional neural networks (CNN) to solve this inverse problem, predicting the underlying 2D PSF $h$, given only the observed data $g$.

In ultrasound imaging, the system PSF $h$ is spatially variant due to electronic focusing of the acoustic signal. Moreover, distortions caused by phase aberration vary across different spatial locations. However, the PSF with phase aberration is slow varying at each spatial location. Therefore, we assume in this work that the PSFs are locally spatially invariant, meaning the PSF remains constant within a given imaging patch. As the network predicts a single PSF for each patch, it effectively estimates the averaged phase-aberrated PSF within that patch.

In addition, we assume that the scatterer $f$ in equation \ref{eqn:convolution_model} is uniformly random distributed to reduce the ill-posed nature of the problem. Therefore, the network is specifically designed and trained to process only smooth, homogeneous speckle patches as input.

\subsection{Synthetic Data Generation}

We train the CNN using synthetic phase-aberrated PSFs due to the impracticality of acquiring paired phase aberration data at scale.

\begin{figure}[!ht]
\centering
\includegraphics[width=0.8\linewidth]{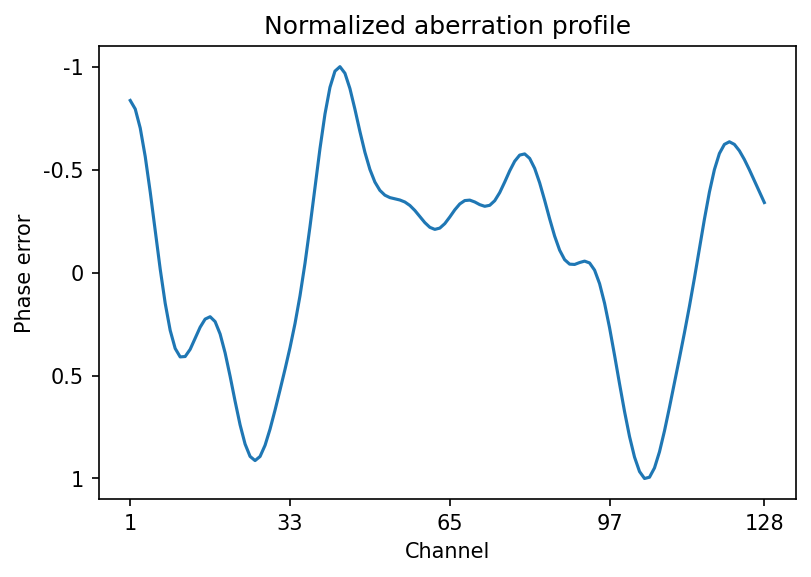}
\caption{An example of a normalized phase aberration profile used for simulating phase aberrated PSF. In this work, we use phase-aberration profiles with correlation length of 5mm. The maximum phase error is adjustable, and ranges from $0\pi$ to $\pi/2$ relative to the center frequency in this work.}
\label{fig:aberration_profile}
\end{figure}

To simulate phase-aberrated PSFs, we employ the near-field phase screen model \cite{flax1988phase}. This model describes the phase aberration process by simulating a phase screen positioned in front of the ultrasound transducer. The phase screen introduces time delays (altering the phase) on the acoustic signals transmitted and received through each channel. Consequently, two-way phase aberration affects both transmit and receive phases. The applied time delay on each channel constitutes the phase aberration profile, as illustrated in Figure \ref{fig:aberration_profile}. In this work, we use phase-aberration profiles with correlation length of 5 mm. All data are simulated by the Field II simulation toolbox \cite{jensen1996field}.


We simulate PSFs using the synthetic transmit aperture (STA) beamforming technique. This approach dynamically focuses the transmitted and received acoustic waves, ensuring that PSFs generated with STA are well-focused at every imaging depth. By eliminating focusing errors caused by beamforming in the simulated PSFs, the network can specifically address errors arising from phase aberration. The parameters used for the simulation are outlined in Table \ref{table:simulation_experiment_parameters}. PSFs were simulated at various transmit center frequencies and imaging depths to ensure the trained network can effectively infer PSFs for different transducers and imaging conditions. A total of 2000 data pairs were generated, with 1800 pairs allocated for training and 200 pairs for validation.


During the training phase of the network, each synthetic PSF is convolved with a uniformly distributed scatterer map to generate homogeneous speckle patches. These scatterer maps are dynamically generated (on-the-fly) for each training step to ensure the generated speckle patches are as diverse as possible. This approach enables the network to learn the generalized concept of phase aberration independently of specific scatterer maps. 

\begin{figure}[!ht]
\centering
\includegraphics[width=1\linewidth]{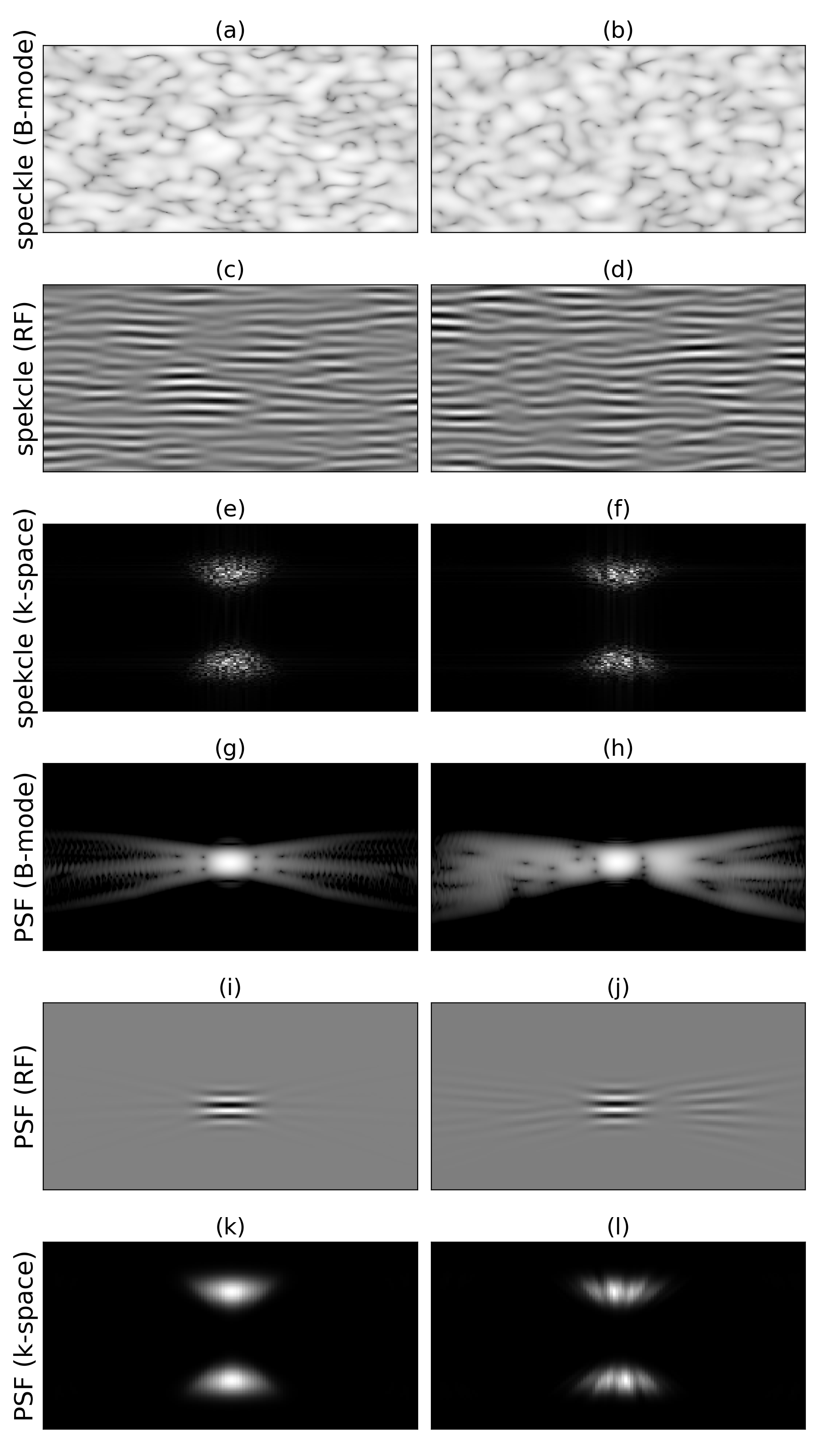}
\caption{Two examples of training data. The left column is a training data without phase aberration, and the right column is a data with phase aberration. The first three rows show the speckle data used as the network's input, visualized in B-mode, RF, and k-space forms, respectively. The last three rows show the PSFs used as the network's target, also visualized in B-mode, RF, and k-space forms, respectively. All B-mode data are displayed with a dynamic range of 60 dB, and the complex k-space data are visualized as their power spectra.}
\label{fig:simulated_data}
\end{figure}

Figure \ref{fig:simulated_data} presents two examples of training data in B-mode, RF, and k-space domains. We can see asymmetric amplitude and increased sidelobe energy caused by phase aberration in Figure \ref{fig:simulated_data}(h). Since the scatterers are uniformly random distributed, the speckle in k-space domains is the pixel-wise multiplication of the PSF's frequency spectrum and the scatterer's frequency spectrum. This results in Figure \ref{fig:simulated_data}(e) appearing as a noisy version of Figure \ref{fig:simulated_data}(k) in their respective power spectra.

A training pair consists of a speckle region and its corresponding underlying PSF. For RF data, Figure \ref{fig:simulated_data}(c) and (i) form a training pair. Similarly, for k-space data, Figure \ref{fig:simulated_data}(e) and (k) form a training pair.

The size of the input speckle patch is chosen to be large enough to include the entire mainlobe and sidelobe of the PSF. The beamspacing parameters used for simulating PSF and speckle patch are shown in table \ref{tab:beamspacing_parameters}.

\begin{table}[h]
\caption{Parameters for Aberrated PSF Generation}
\centering
\begin{tabular}{@{}cc@{}}
\toprule
\textbf{Parameters} & \textbf{} \\
\midrule
Number of transducer elements & 128 \\
Sample frequency & 100 MHz \\
Center frequency & 3 - 7.5 MHz \\
Acoustic speed & 1540 m/s \\
Probe fractional bandwidth & 0.3 - 0.8 \\
Pitch & 0.3 mm \\
Height & 5mm \\
Depth & 10 - 40 mm \\
f\# & 2 \\
Max phase error & 0 - $\pi$ / 2 \\ 
Aberration correlation length & 5 mm \\
\bottomrule
\end{tabular}
\label{table:simulation_experiment_parameters}
\end{table}

\begin{table}[h]
\caption{Beamspacing for PSF Simulation}
\centering
\begin{tabular}{@{}cc@{}}
\toprule
\textbf{Parameters} & \textbf{} \\
\midrule
Patch resolution & 256 x 256 (pixel) \\
Patch height & 16 $\lambda$ \\
Axial sample interval & $\lambda/16$ \\
Patch width & 32 $\lambda$ \\
Lateral sample interval & $\lambda/32$ \\
\bottomrule
\end{tabular}
\label{tab:beamspacing_parameters}
\end{table}

\subsection{Convolutional Neural Network}

We employ image-to-image convolutional neural networks (CNNs) to estimate the 2D phase-aberrated PSF. Within this framework, the network performs a form of blind deconvolution on the scatterer distribution to reconstruct the system's phase-aberrated PSF. A U-Net is designed to process patches of beamformed radio-frequency (RF) speckle data, predicting the corresponding phase-aberrated RF PSF for each patch \cite{unet}. Additionally, a complex U-Net is developed to operate on patches of beamformed k-space speckle data, predicting the phase-aberrated k-space PSF \cite{trabelsi2018deep}. The complex U-Net separates the real part and imaginary part into two channels during the convolution layer.

In the k-space domain, the network's functionality is similar to denoising or super-resolution applied to the power spectrum, mapping Figure \ref{fig:simulated_data}(e) to Figure \ref{fig:simulated_data}(k). Therefore, this relationship is more direct and intuitive compared to the RF domain, where the network maps Figure \ref{fig:simulated_data}(c) to Figure \ref{fig:simulated_data}(i). The results demonstrate that the complex U-Net operating in the k-space domain achieves superior performance in estimating phase-aberrated PSFs.

\subsection{Sidelobe-Emphasized Loss Function}

The training of neural networks relies on designated loss functions that guide the optimization process. For estimating PSFs, the loss function should quantify the error (i.e., difference) between the target PSF and the predicted PSF.

Traditional pixel-wise loss functions like L1 or L2 loss are inadequate for measuring the similarity between PSFs for two primary reasons. Firstly, the signal intensity within the sidelobe region of PSFs is significantly lower than in the mainlobe region. Consequently, a pixel-wise loss function would mainly emphasize errors in the mainlobe, potentially leading to less effective optimization guidance for the PSF's sidelobe region. Secondly, the target PSF and the predicted PSF may not align perfectly within the patch, causing slight spatial shifts between them. Since pixel-wise losses are not shift invariant, the target PSF's alignment could vary, potentially resulting in blurry PSF predictions.

In contrast, SSIM loss focuses on the overall structures of the image and is calculated in a patch-wise manner. The patch-wise approach may better encourage the generation of sidelobes in the predicted PSFs, improving their overall structural accuracy.

In this work, we utilize perceptual loss that calculates on log-compressed B-mode PSFs to guide the network's training. Perceptual loss is computed as the L2 norm between latent features extracted from a VGG19 network pretrained on ImageNet \cite{perceptual_loss,vgg,imagenet}. Specifically, we extract latent features from the 2nd, 5th, 8th, and 13th layers of the VGG19 network.



For the U-Net that infers RF PSFs, the PSF is first baseband demodulated and log-compressed to convert it into B-mode PSF before calculating the perceptual loss and its gradient. The loss function is defined as,

\begin{equation}
L_{RF}(\widehat{y}, y)=\left \| \sum_{i\epsilon N}^{} \phi_i(log(b(\widehat{y})))-\phi_i(log(b(y))) \right \|^2_2
\label{eqn:loss_function_rf}
\end{equation}

\noindent
where $\widehat{y}$ and $y$ are the predicted and target PSFs, respectively. $N$ denotes the selected output layers of the VGG19 network for perceptual loss. $\phi_i$ represents the latent feature from the i-th layer of the VGG19 network. $\left \| \cdot \right \|_2^2$ denotes the L2 norm. $b(\cdot)$ is the baseband demodulation function, and $log(\cdot)$ is the log compression function with a gain and dynamic range of 60 dB.

For the complex U-Net that infers k-space PSFs, the k-space PSF is first transformed back to the RF domain using the 2D inverse Fourier Transform. Subsequently, baseband demodulation and log compression are applied to convert the PSF into the B-mode PSF for perceptual loss calculation and gradient computation. The loss function is defined as, 

\begin{equation}
L_{K}(\widehat{y}, y)=\left \| \sum_{i\epsilon N}^{} \phi_i(log(b(ifft(\widehat{y}))))-\phi_i(log(b(ifft(y)))) \right \|^2_2 
\label{eqn:loss_function_kspace}
\end{equation}

\noindent
where $ifft(\cdot)$ indicates the 2D inverse Fourier Transform.

\section{Results}




\begin{figure*}[!ht]
\centering
\includegraphics[width=1.0\linewidth]{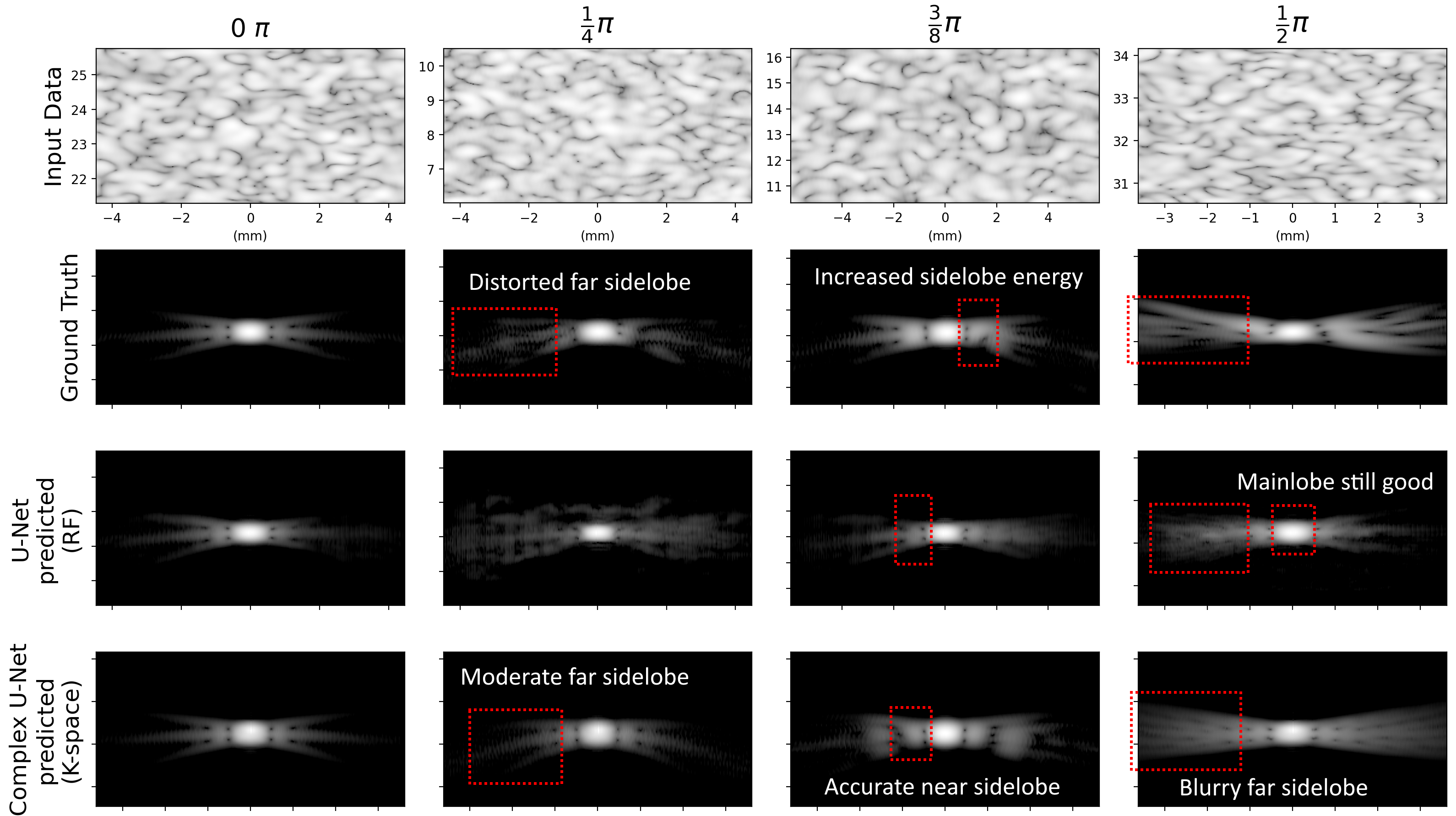}
\caption{Results of the simulation. The first row is the input data, second row is the target ground truth PSFs, the third row is the PSFs predicted by U-Net, and the fourth row is the PSFs predicted by complex U-Net. The phase aberration levels for each column from left to right is 0 (no aberration), $\frac{1}{4}\pi$, $\frac{3}{8}\pi$, and $\frac{1}{2}\pi$, respectively. Note that the PSFs are inferred in RF and k-space domain for U-Net and complex U-Net, respectively, and are displayed in log-scale only for visualization purpose. The images are displayed in log-scale with a dynamic range of 60 dB.}
\label{fig:predicted_PSF_by_unet3}
\end{figure*}

\subsection{Objective Metrics}

Structural similarity index (SSIM) is employed to evaluate the performance of the trained network \cite{SSIM}, and is calculated on log-compressed PSFs with a dynamic range of 60 dB. A well-estimated PSF exhibits high SSIM values, indicating a high structural similarity between the predicted and ground truth PSFs. Additionally, we introduce the lateral beam pattern difference (LBPD) metric to evaluate the performance of the predicted PSF's sidelobes, which are primarily affected by phase aberration. LBPD is computed as the L2 distance between the projected lateral beam patterns of the ground truth PSF and the predicted PSFs. The projected lateral beam pattern is derived as the intensity projection along the z-axis of a log-compressed PSF. A well-focused PSF with minimal aberration typically exhibits the majority of its energy concentrated in the mainlobe, whereas a highly aberrated PSF shows increased intensity in the sidelobes. Therefore, well-predicted PSFs should yield low LBPD values.



To evaluate the prediction accuracy of the PSF across different regions, we employ the intersection over union (IoU) score proposed by \cite{yalin_revised_lrelu}. The IoU score assesses the log-compressed PSF across three distinct sections: the mainlobe, near sidelobe, and far sidelobe, which are defined based on their intensity levels. Following the methodology presented in \cite{yalin_revised_lrelu}, the PSF intensity is divided into 20 dB intervals. Specifically, the mainlobe section corresponds to intensity levels between 40 and 60 dB (denoted as IoU1), the near sidelobe section includes levels between 20 and 40 dB (denoted as IoU2), and the far sidelobe section covers levels between 0 and 20 dB (denoted as IoU3). The overall IoU score is computed as the average of the IoU scores across these three sections.


\subsection{Simulation Experiment}

To assess the effectiveness of the proposed method, we evaluated the performance of U-Net and complex U-Net in accurately estimating phase-aberrated PSFs using synthetic data.

The U-Net is trained with speckle data and PSFs in RF domain, and the complex U-Net is trained with speckle data and PSFs in complex k-space domain. Training utilizes the Adam optimizer with an exponential decay learning rate, starting at 1e-3 and reducing by a factor of 0.1 every 10 epochs. Training spans 100 epochs with a batch size of 1. All networks employ LeakyReLU activation functions for all layers. The implementation is carried out in TensorFlow, using a Nvidia RTX2080Ti GPU for training.

Table \ref{tab:simulation_experiment_performances} presents the objective performance metrics of the networks. The metrics are the average calculated on the 200 pairs of validation data. The results show that complex U-Net outperforms U-Net, indicating superior performance for estimating PSFs in the k-space domain. Figure \ref{fig:predicted_PSF_by_unet3} displays four samples of PSFs predicted by U-Net and complex U-Net with different aberration levels, together with their corresponding ground truth. 

When there is no aberration, the network accurately predicts the entire PSF, including both the mainlobe and sidelobe regions. However, as the aberration level increases, the networks provide accurate estimations primarily in the mainlobe and near-sidelobe regions, while the predictions in the far-sidelobe regions become blurry. This degradation may be attributed to the high distortion caused by phase aberration and the ill-posed nature of PSF estimation, which makes it challenging for the networks to learn these regions effectively.

\begin{table}[h]
\caption{Simulation Results}
\centering
\begin{tabular}{@{}cccc@{}}
\toprule
 & SSIM ($\uparrow$) & LBPD ($\downarrow$) & IoU ($\uparrow$) \\
\midrule
U-Net  & 0.807 & 72.79 & 0.617 \\
Complex U-Net  & \textbf{0.861} & \textbf{35.82} & \textbf{0.713} \\
\bottomrule
\end{tabular}
\label{tab:simulation_experiment_performances}
\end{table}

\subsection{Aberration Limitation}

In the previous section, we observed a decline in prediction performance as the aberration level increased. This outcome is expected, as higher aberration levels lead to greater distortion and incoherence in PSFs, which in turn increases the sidelobe energy. Consequently, predicting the sidelobe regions of the PSFs becomes more challenging for the network under high phase aberration levels.

In this section, we evaluate the network’s ability to infer PSFs under varying levels of phase aberration. Table \ref{tab:simulation_aberration_metrics} presents the objective metrics used to evaluate the network's performance in estimating PSFs for different aberration levels. The metrics are the average calculated on the 200 pairs of validation data. 

The results indicate a decline in estimation accuracy as aberration levels increase, particularly affecting the accuracy of the sidelobe regions of the PSFs (i.e., IoU2 and IoU3). In contrast, the performance for the mainlobe regions (i.e., IoU1) remains relatively consistent. This decline is attributed to the high distortion of the PSF's far sidelobe under elevated phase aberration levels, which challenges the network's ability to accurately predict this region.


\begin{table}[h]
\small
\caption{PSF Estimation Performance of Different Aberration Levels}
\centering
\begin{tabular}{@{}cccccc@{}}
\toprule
 & SSIM ($\uparrow$) & IoU1 ($\uparrow$) & IoU2 ($\uparrow$) & IoU3 ($\uparrow$) & LBPD ($\downarrow$) \\
\midrule
$0 \pi$     & \textbf{0.950} & \textbf{0.963} & \textbf{0.774} & \textbf{0.845} & \textbf{5.73} \\
$(1/4) \pi$ & 0.843 & 0.917 & 0.590 & 0.501 & 28.13         \\
$(3/8) \pi$ & 0.837 & 0.929 & 0.603 & 0.481 & 40.25         \\
$(1/2) \pi$ & 0.815 & 0.932 & 0.587 & 0.429 & 69.16        \\
\bottomrule
\end{tabular}
\label{tab:simulation_aberration_metrics}
\end{table}



\subsection{Loss Functions}

In this section, we evaluate the effectiveness of various loss functions for training the network, including L1 loss, L2 loss, SSIM loss, and perceptual loss. Additionally, we assess the performance of these loss functions when applied to RF data, complex k-space data, and log-compressed B-mode data. 

\begin{figure}[!ht]
\centering
\includegraphics[width=1\linewidth]{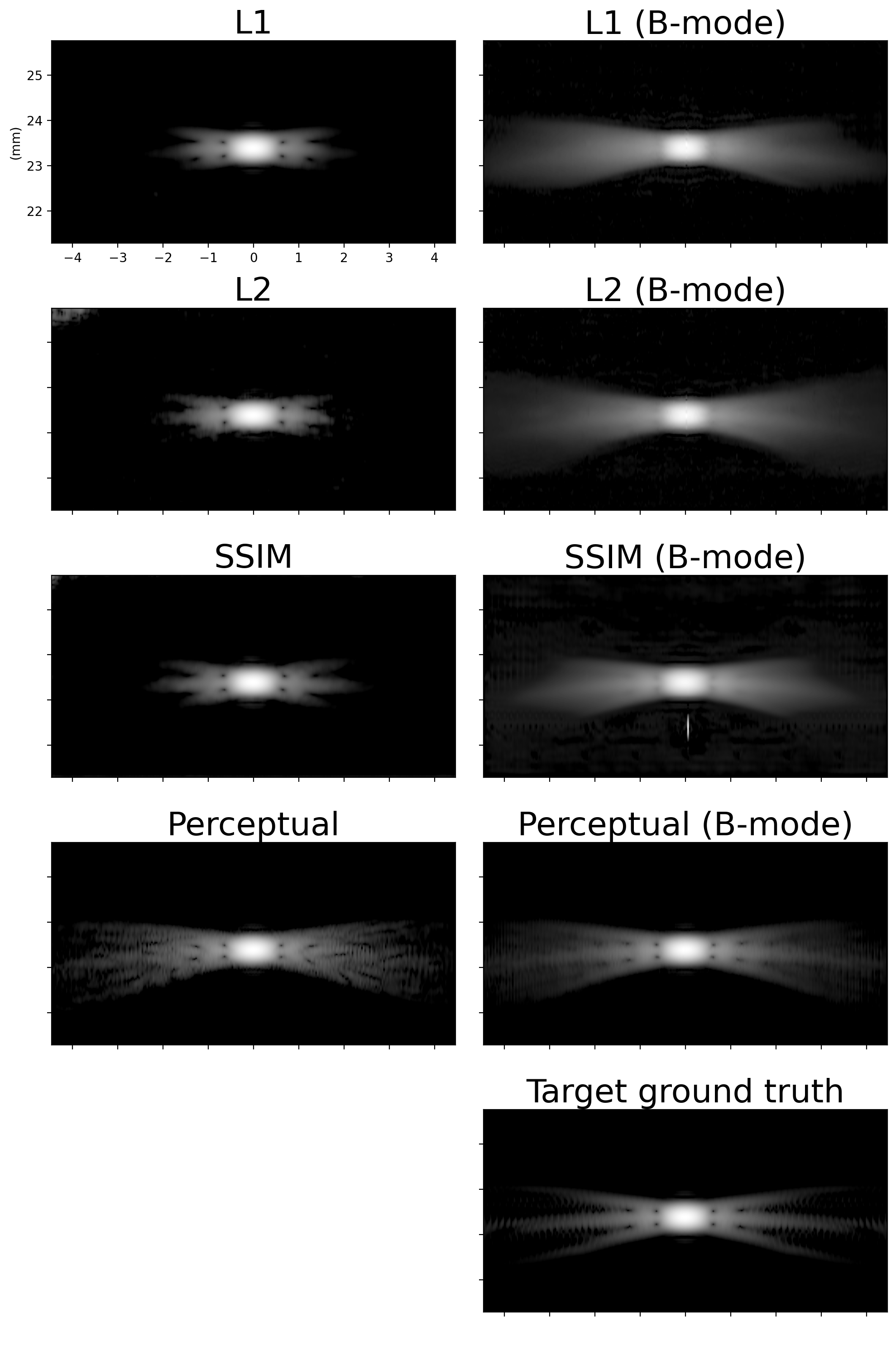}
\caption{The PSFs predicted by U-Net trained with different loss functions and the target ground truth PSF. The title of each image indicates the loss function used to train the U-Net, with "(B-mode)" specifying that the loss function was calculated on B-mode images. The input speckle data and PSFs are processed in the RF form, whereas the displayed log-compressed images are shown only for visualization purpose. All images are displayed in log-scale with a dynamic range of 60 dB.}
\label{fig:results_loss_rf}
\end{figure}

Figure \ref{fig:results_loss_rf} and Table \ref{tab:loss_function_performances} present the predicted PSF results from U-Net operating in the RF domain, trained with various loss functions. Pixel-wise loss functions, such as L1 and L2 loss, effectively capture the mainlobe but struggle with accurate prediction in the sidelobe regions due to the dominance of mainlobe intensity, as reflected in their poor LBPD scores. Similarly, SSIM loss faces the same limitation. Loss functions computed on log-compressed B-mode data better motivate sidelobe generation; however, the generated sidelobes appear blurry. In contrast, perceptual loss produces a clear representation of both the mainlobe and sidelobes.

\begin{figure}[!ht]
\centering
\includegraphics[width=1\linewidth]{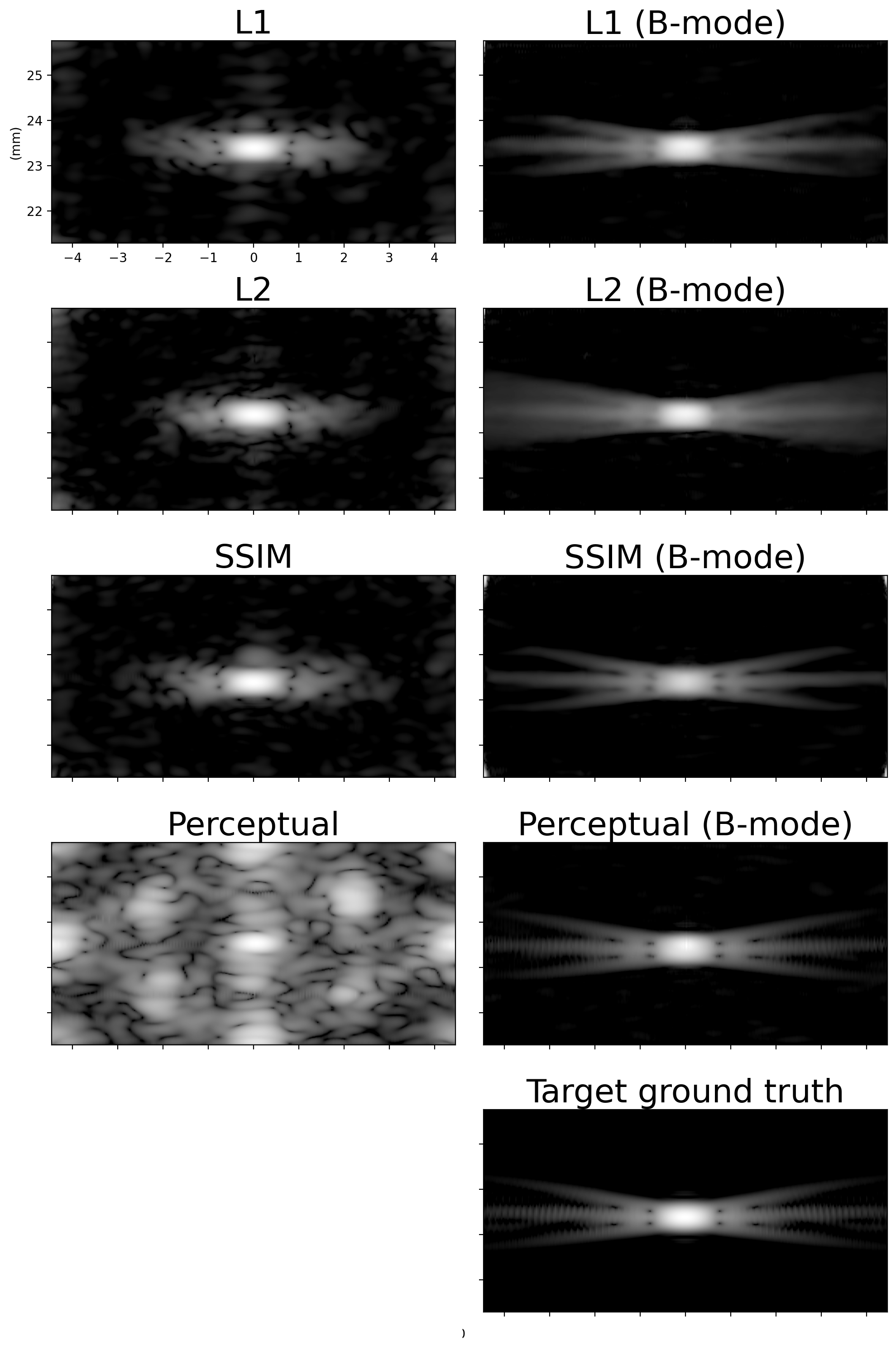}
\caption{The PSFs predicted by complex U-Net trained with different loss functions and the target ground truth PSF. The title of each image indicates the loss function used for training, with "(B-mode)" specifying that the loss function was calculated on B-mode images. The input speckle data and PSFs are processed in complex k-space data, and the displayed log-compressed images are shown only for visualization purpose. All images are displayed in log-scale with a dynamic range of 60 dB.}
\label{fig:results_loss_k}
\end{figure}

Figure \ref{fig:results_loss_k} shows the results from complex U-Net operating in the k-space domain. L1, L2, and SSIM losses perform well only in the mainlobe region and fail to generate meaningful sidelobes for the PSF. When perceptual loss is applied directly in the k-space domain, it separates the real and imaginary parts of the complex data, disrupting the inherent correlations and relationships within the k-space data, ultimately resulting in a complete failure to generate PSFs. Loss functions computed on log-compressed B-mode data generally yield better results, likely because B-mode data more closely resembles natural images compared to RF or k-space data. This resemblance enhances the effectiveness of conventional loss functions, particularly perceptual loss trained on natural images from ImageNet, which achieves the best LBPD and IoU scores.



\begin{table}[h]
\caption{Performance from Different Loss Functions}
\centering
\begin{tabular}{@{}cccccc@{}}
\toprule
 & SSIM ($\uparrow$) & LBPD ($\downarrow$) & IoU ($\uparrow$) \\
\midrule
U-Net (RF) \\
\midrule
L1                  & 0.784 & 905.64 & 0.446          \\
L2                  & 0.759 & 266.34 & 0.426          \\
SSIM                & 0.784 & 183.16 & 0.452          \\
Perceptual          & 0.791 & 74.43  & 0.595          \\
L1 (B-mode)         & 0.812 & 53.52  & 0.616          \\
L2 (B-mode)         & 0.760 & 58.48  & 0.577          \\
SSIM (B-mode)       & 0.779 & 63.64  & 0.550          \\
Perceptual (B-mode) & 0.807 & 72.79  & 0.617          \\
\midrule
Complex U-Net (k-space) \\
\midrule
L1                  & 0.678 & 91.61 & 0.477          \\
L2                  & 0.645 & 110.89 & 0.439          \\
SSIM                & 0.627 & 67.02 & 0.454          \\
Perceptual          & 0.039 & 461.85  & 0.059          \\
L1 (B-mode)         & 0.863 & 38.75  & 0.712          \\
L2 (B-mode)         & 0.856 & 38.43  & 0.706          \\
SSIM (B-mode)       & \textbf{0.866} & 69.47  & 0.712          \\
Perceptual (B-mode) & 0.861 & \textbf{35.82}  & \textbf{0.713}          \\
\bottomrule
\end{tabular}
\label{tab:loss_function_performances}
\end{table}

\section{Discussion}
\label{section:discussion}

\subsection{Limitations}

The results indicate that the network struggles to accurately estimate the PSF's far sidelobe region, especially under strong phase aberration. This limitation may be due to the inherently ill-posed nature of the inverse problem, where the PSF's far sidelobe region is highly sensitive to phase aberration and dominated by low-intensity signals. Additionally, the far sidelobe regions exhibit complex spatial characteristics, including asymmetric amplitudes and distorted structures, which make them particularly challenging for the network to learn effectively.

Given that our synthetic training pairs consist of homogeneous speckled regions, it's anticipated that these inputs are devoid of any structural features. However, in real-world scenarios, it can be challenging to locate sufficiently large homogeneous patches suitable for the network's input. One approach to address this challenge is to incorporate nonhomogeneous patches with diverse structures into the training data. This strategy encourages the network to learn to disregard these structures and accurately predict PSFs, at the expense of increased network complexity and training difficulty.

Another solution is to use feature detection algorithms to avoid the selection of nonhomogeneous region \cite{shen2022multi,wang2008novel,slabaugh2006information}. These algorithms can initially detect features, allowing for the subsequent selection of non-feature homogeneous patches as suitable inputs for the network. In cases where the identified homogeneous patch is too small for the network's requirements, mirror padding can effectively enlarge the patch to the desired size.



\subsection{Two-stage Approach}

The primary objective of estimating PSFs in ultrasound imaging is to integrate them with other techniques such as deconvolution, speckle reduction, or phase aberration correction. A practical approach to achieve this is through a two-stage method, where the network is integrated into the processing pipeline. Initially, in the first stage, the network estimates the phase aberrated PSFs, which are subsequently utilized to optimize the performance of other algorithms in the second stage of the pipeline.

The rationale behind opting for a two-stage approach instead of directly applying an end-to-end deep network for tasks like deconvolution, speckle reduction, or phase aberration correction in ultrasound imaging lies in the advantages that it offers tunability and interpretability. In medical imaging, where fidelity and robustness are paramount, directly applying a deep network can inadvertently introduce fake artifacts or obscure critical features, potentially leading to incorrect diagnoses.

Conversely, in a two-stage approach, any inaccuracies in the estimated PSF primarily impact subsequent algorithms, allowing for the preservation of original image features. This methodology, which is also prevalent in other imaging fields \cite{tseng2021differentiable,frosio2015machine,tseng2019hyperparameter}, is considered more suitable for medical imaging applications.

\subsection{Execution Speed}

A single inference for U-Net and complex U-Net takes approximately 16.1 ms and 62.2 ms, respectively, on an RTX2080Ti GPU. These correspond to frame rates of 62 and 16 frames per second, respectively, meeting the real-time requirement of 15 frames per second for ultrasound B-mode imaging.

Ideally, the network should infer the phase-aberrated PSF at the focusing depth to isolate the focusing error caused solely by phase aberraiton. For scenarios with multiple focus points, multiple GPUs can be employed in parallel to infer the phase-aberrated PSFs at each location, ensuring efficient and real-time processing across all focal regions.



\section{Conclusions}
\label{section:conclusions}

In this study, we employ deep learning techniques to train a network for estimating phase aberrated PSFs in ultrasound imaging. The network is trained using synthetic phase aberration data generated according to the near-field phase screen model. Additionally, a novel loss function is utilized to effectively guide the network's training process. The results demonstrate that the complex K-space network outperforms the RF one, and the network performs well in estimating PSFs even in the presence of phase aberration.



\bibliographystyle{IEEEtran}
\bibliography{submission/main}



\end{document}